\documentclass[aps,pra,twocolumn,superscriptaddress,showpacs]{revtex4-1}
\def\be{\begin{equation}}
\def\ee{\end{equation}}
\def\bea{\begin{eqnarray}}          
\def\eea{\end{eqnarray}}
\def\bi{\begin{itemize}}
\def\ei{\end{itemize}}

\usepackage{graphicx}

\begin{document}

\title{ 
             Space-time renormalization in phase transition dynamics
}

\author{Anna Francuz} 
\affiliation{Instytut Fizyki Uniwersytetu Jagiello\'nskiego,
             ul. {\L}ojasiewicza 11, 30-348 Krak\'ow, Poland}
\affiliation{Theoretical Division, LANL, Los Alamos, New Mexico 87545, USA} 

\author{Jacek Dziarmaga} 
\affiliation{Instytut Fizyki Uniwersytetu Jagiello\'nskiego,
             ul. {\L}ojasiewicza 11, 30-348 Krak\'ow, Poland}
             
\author{Bart{\l}omiej Gardas}
\affiliation{Theoretical Division, LANL, Los Alamos, New Mexico 87545, USA}             
\affiliation{Institute of Physics, University of Silesia, 40-007 Katowice, Poland}             
             
\author{Wojciech H. Zurek}
\affiliation{Theoretical Division, LANL, Los Alamos, New Mexico 87545, USA}

\date{ December 13, 2015 }


\begin{abstract}
When a system is driven across a quantum critical point at a constant rate its evolution must become non-adiabatic as the relaxation time $\tau$ 
diverges at the critical point. According to the Kibble-Zurek mechanism (KZM), the emerging post-transition excited state is characterized by a finite correlation length $\hat\xi$ set at the time $\hat t=\hat \tau$ when the critical slowing down makes it impossible for the system to relax to the equilibrium defined by changing parameters. This observation naturally suggests a dynamical scaling similar to renormalization familiar from the equilibrium critical phenomena.
We provide evidence for such KZM-inspired spatiotemporal scaling by investigating an exact solution of the transverse field quantum Ising chain in the 
thermodynamic limit.  
\end{abstract}


\maketitle

\section{ Introduction } 

The study of the dynamics of second-order phase transitions started in the cosmological setting with the observation by 
Kibble \cite{Kibble76,Kibble80} that, in course of the rapid cooling that follows Big Bang, distinct domains of the nascent Universe 
will be forced to choose broken symmetry vacua independently. 
Their incompatibility will typically lead to topological defects that may have observable consequences.

The relativistic causal horizon is no longer a useful constraint in condensed matter settings, 
but one can still define a sonic horizon that plays a similar role \cite{Zurek85,Zurek93,Zurek96}. 
The usual estimate of the sonic horizon relies on the scaling of the relaxation time and of the healing length that depend on the dynamical and spatial critical exponents $z$ and $\nu$ characteristic for the relevant universality class.
The estimate predicts a characteristic time-scale
$
\hat t \sim \tau_Q^{z\nu/(1+z\nu)}\sim \hat \tau
$
and a correlation length (length-scale)
$
\hat\xi \sim \tau_Q^{\nu/(1+z\nu)},
$
where the quench time $\tau_Q$ quantifies the rate of the transition. The correlation length enables prediction of the scaling 
exponent that governs the number of the generated excitations (e.g., the density of topological defects, when the relevant homotopy group allows for their formation) as a function of $\tau_Q$ for a wide range of quench rates.

The Kibble-Zurek mechanism has been confirmed by numerical simulations \cite{LagunaZ1,YZ,DLZ99,ABZ99,ABZ00,BZDA00,ions20,ions2,WDGR11,dkzm1,dkzm2,Nigmatullin11,DSZ12,holo} 
and, to a lesser degree, and with more caveats, by experiments \cite{Chuang91,Bowick94,Ruutu96,Bauerle96,Carmi00,Monaco02,Monaco09,Maniv03,Sadler06,Golubchik10,Chae12,Griffin12,Schaetz13,EH13,Ulm13,Tanja13,Anderson08,
Lamporesi13} in a variety of settings, with most recent results in solid state physics as well as in gaseous Bose-Einstein condensates providing suggestive evidence of KZM scalings \cite{Chae12,Griffin12,DalibardSupercurrents,DalibardCoherence,ferroelectrics,Hadzibabic}. 

Refinements and extensions of KZM include phase transition in inhomogeneous systems (see \cite{DKZ13} for recent overview) and applications that go beyond topological defect creation (see e.g. \cite{DQZ11,Zurek09,DZ10,Cincio}). Recent reviews related to KZ mechanism are also available \cite{Kibble03,Kibble07,Dziarmaga10,Polkovnikov11,DZ13}.

We consider a zero-temperature quantum phase transition in the transverse-field quantum Ising chain. 
Despite important differences with respect to thermodynamic phase transitions -- where thermal rather 
than quantum fluctuations act as seeds of symmetry breaking -- the KZM can be generalized to quantum phase transitions
\cite{Bishop,Damski2005,Dorner2005,Dziarmaga2005,Polkovnikov2005,ind}, see also \cite{Dziarmaga10,Polkovnikov11,DZ13} for reviews.
The quantum regime was also addressed in some of the recent experiments \cite{Esslinger,deMarcoclean,Schaetz,deMarcodisorder,chinskiLZ}.

In this paper we propose what can be considered a generalization and extension of the predictive power of KZM: In the adiabatic limit, when $\tau_Q\to\infty$, both $\hat t$ and $\hat \xi$ diverge. Hence, one can expect that they should be the only relevant time and length scales in the low frequency and long wavelength regime. This in turn suggests a dynamical scaling hypothesis, similar to the one that underlies renormalization paradigm that is so useful for the equilibrium phase transitions,
that during the quench all physical observables depend on time $t$ through the rescaled time $t/\hat t$ and on a distance $x$ 
through the rescaled distance $x/\hat\xi$. Though the basic ingredients of the hypothesis were present in the KZM 
from the beginning (see e.g. discussion of the re-scaling of Gross-Pitaevskii equation in \cite{Zurek96}, as well as \cite{Cincio,ViolaOrtiz,DamskiZurek,DzRams}), its fully fledged form, taking into account the scaling dimension,
was articulated first in \cite{Kolodrubetz} for the correlation function of the ferromagnetic order parameter in the
quantum Ising chain. The idea was developed further in \cite{princeton}.
  
Our aim here is a comprehensive study of this spacetime renormalization-like scaling in the exactly solvable Ising chain. 
We begin with a general discussion of the quantum KZM in section \ref{sec:QKZ}.
It is followed by the statement of the KZM scaling hypothesis in section \ref{sec:scaling}. 
In section \ref{sec:horizon} we discuss the sonic horizon. 
The Ising model is solved in sections \ref{sec:Ising} and \ref{sec:LZ} by mapping to a set of independent Landau-Zener (LZ) systems. 
The scaling in the LZ context is identified in section \ref{sec:LZscaling}. 
Then the same scaling is found in quadratic fermionic correlators \ref{sec:scalingfermions}, 
energy and quasiparticle density \ref{sec:scalingenergy}, spin-spin correlators \ref{sec:scalingCR}, mutual information \ref{sec:scalingmutual}, 
quantum discord \ref{sec:scalingdiscord}, entropy of entanglement \ref{sec:scalingentropy}, and entanglement gap \ref{sec:scalinggap}. 
We conclude in section \ref{sec:conclusion}. 

\begin{figure}[t]
\includegraphics[width=1.0\columnwidth,clip=true]{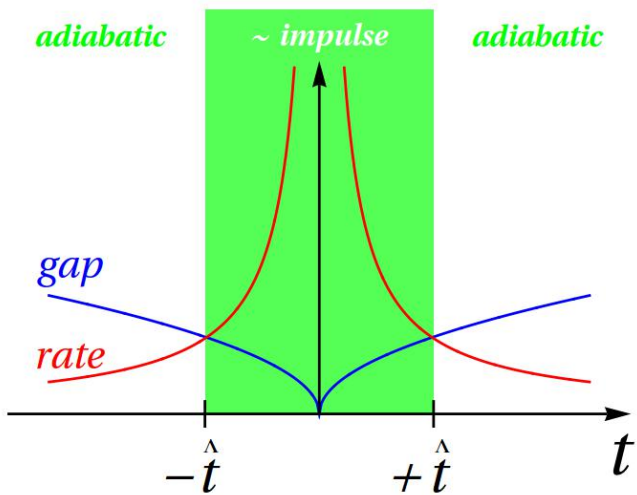}
\caption{  
Schematic illustration of the quantum Kibble-Zurek mechanism.
The Hamiltonian is driven by a linear quench (\ref{quench}) across the critical point at $t=0$. 
The energy gap (\ref{Delta}) closes like $|\epsilon|^{z\nu}$, 
where $z$ and $\nu$ are universal exponents, 
and the transition rate (\ref{rate}) diverges at the critical point,
hence the evolution cannot be adiabatic between $-\hat t$ and $+\hat t$.
The above figure shows the gap for $z=1$ and $\nu=1/2$, i.e., the mean field scalings.
In the quantum Ising model $z=\nu=1$ and the gap closes linearly as $t\to0$. 
}
\label{figadiabimpulse}
\end{figure}

\section{ Quantum Kibble-Zurek mechanism } \label{sec:QKZ}

A distance from a quantum critical point can be measured with a dimensionless parameter $\epsilon$. The ground state of the Hamiltonian $H(\epsilon)$ changes character (e.g., breaks a symmetry) when $\epsilon=0$. Thus, $\epsilon$ plays a role analogous to the relative temperature in thermodynamic phase transitions. 

The correlation length in its ground state diverges like
\be 
\xi \sim |\epsilon|^{-\nu}
\ee
and the relevant gap closes, 
\be 
\Delta \sim |\epsilon|^{z\nu},
\label{Delta}
\ee
see Figure \ref{figadiabimpulse}.
The system, initially prepared in its ground state, is driven across the critical point by a linear quench,
\be 
\epsilon(t)=\frac{t}{\tau_Q},
\label{quench}
\ee 
with a quench time $\tau_Q$. Nonlinear ``protocols" can be also considered \cite{nonlin,princeton}, 
but we shall not deal with them here. 

The evolution sufficiently far from the critical point is initially adiabatic. However, the rate of change
of epsilon,
\be 
\left|\frac{\dot\epsilon}{\epsilon}\right|=\frac{1}{|t|}, 
\label{rate}
\ee
diverges at the gapless critical point. Therefore, evolution (e.g., of the order parameter) cannot be adiabatic in its neighborhood between $-\hat t$ and $\hat t$,
see Fig. \ref{figadiabimpulse}.
Here $\hat t$ is the time when the gap (\ref{Delta}) equals the rate (\ref{rate}), so that:
\be 
\hat t \sim \tau_Q^{z\nu/(1+z\nu)} \sim \hat \tau.
\ee
Just before the adiabatic-to-non-adiabatic crossover at $-\hat t$, the state of the system is still 
approximately the adiabatic ground state at $\epsilon=-\hat\epsilon$, where
\be 
\hat\epsilon = \frac{\hat t}{\tau_Q} \simeq \tau_Q^{-1/(1+z\nu)},
\ee
with a correlation length
\be 
\hat\xi \sim  \hat\epsilon^{-\nu} \sim \tau_Q^{\nu/(1+z\nu)}.
\ee
In a zeroth-order impulse approximation (which is the ``caricature'' of the KZM often found in papers) this state ``freezes out'' at $-\hat t$ and literally does not change until $\hat t$. 
At $\hat t$ the frozen state is no longer the ground state but an excited state with a correlation length $\hat\xi$.
It is the initial state for the adiabatic process that follows after $\hat t$.

There are cases where this oversimplified view suffices \cite{Damski2005}. Moreover, as we shall see below, it predicts the same scalings for $\hat \xi$ as the original derivation \cite{Zurek85,Zurek96} based on the size of the sonic horizon.

\section{ Space-time renormalization scaling hypothesis } \label{sec:scaling}

No matter how accurate is the impulse approximation or the above ``freeze-out scenario'', 
the scaling argument establishes $\hat\xi$ and $\hat t$, 
interrelated via 
\be 
\hat t\sim\hat\xi^z,
\ee 
as the relevant scales of length and time. 
What is more, in the adiabatic limit, when $\tau_Q\to\infty$, both scales diverge becoming the unique 
scales in the long wavelength and low frequency limit. 
Like in the static critical phenomena, 
this uniqueness implies a scaling hypothesis:
\be 
\langle\psi(t)| O(x) |\psi(t)\rangle = \hat\xi^{-\Delta_O} F_O\left(t/\hat t,x/\hat\xi\right).
\ee
Here $|\psi(t)\rangle$ is the state during the quench, 
$O(x)$ is an operator depending on a distance $x$, 
$F_O$ is its scaling function, 
and $\Delta_O$ its scaling dimension. 
This hypothesis is analogous to the static one in the ground state $|\psi_{\rm GS}\rangle$,
\be 
\langle\psi_{\rm GS}| O(x) |\psi_{\rm GS}\rangle = 
\xi^{-\Delta_O^{\rm GS}} F_O^{\rm (GS)}\left(x/\xi\right),
\ee
where $\xi$ is a diverging correlation length near a quantum critical point. 

The diverging scales, $\hat\xi$ and $\hat t$, become the unique scales in a coarse-grained description
at large distances and long times, but the scaling hypothesis is not warranted to hold at short 
microscopic distances of a few lattice sites, where microscopic scales remain relevant. This
is the same as in the static critical phenomena.  

The analogy to the static case is nearly an identity near $t/\hat t=-1$, 
where $|\psi(t)\rangle=|\psi_{\rm GS}\rangle$ and $\hat\xi=\xi$.
Consequently, 
\bea 
&& 
F_O(-1,x/\hat\xi)=F_O^{\rm (GS)}\left(x/\hat\xi\right), \\
&&
\Delta_O=\Delta_O^{\rm (GS)}.
\eea 
The dynamical dimension is the same as the static one.
Exploiting further the adiabaticity before $t/\hat t=-1$, 
the adiabatic scaling function is well approximated by
\bea
F_O\left( t/\hat t<-1 , x/\hat\xi \right) &=&
\left(\xi/\hat\xi\right)^{-\Delta_O}
F_O^{\rm (GS)}\left(\frac{x/\hat\xi}{\xi/\hat\xi}\right) \nonumber\\
&=&
\left(t/\hat t\right)^{\nu\Delta_O}
F_O^{\rm (GS)}\left(\frac{x/\hat\xi}{(t/\hat t)^{-\nu}}\right).
\eea
Here $\xi$ is the correlation length in the adiabatic ground state before $-\hat t$.
It depends on time like 
$
\xi/\hat\xi=
\left(\epsilon/\hat\epsilon\right)^{-\nu}=
\left(t/\hat t\right)^{-\nu}
$.
What is more, in the impulse approximation, 
the non-adiabatic scaling function should not depend on the rescaled time:
\bea 
F_O\left( -1<t/\hat t<1 , x/\hat\xi \right) &=&
F_O\left(-1,x/\hat\xi \right) \nonumber\\
&=&
F_O^{(GS)}\left(x/\hat\xi\right).
\eea
If accurate, the dynamical function would be completely expressible by the static one.

\begin{figure}[t]
\includegraphics[width=0.99\columnwidth,clip=true]{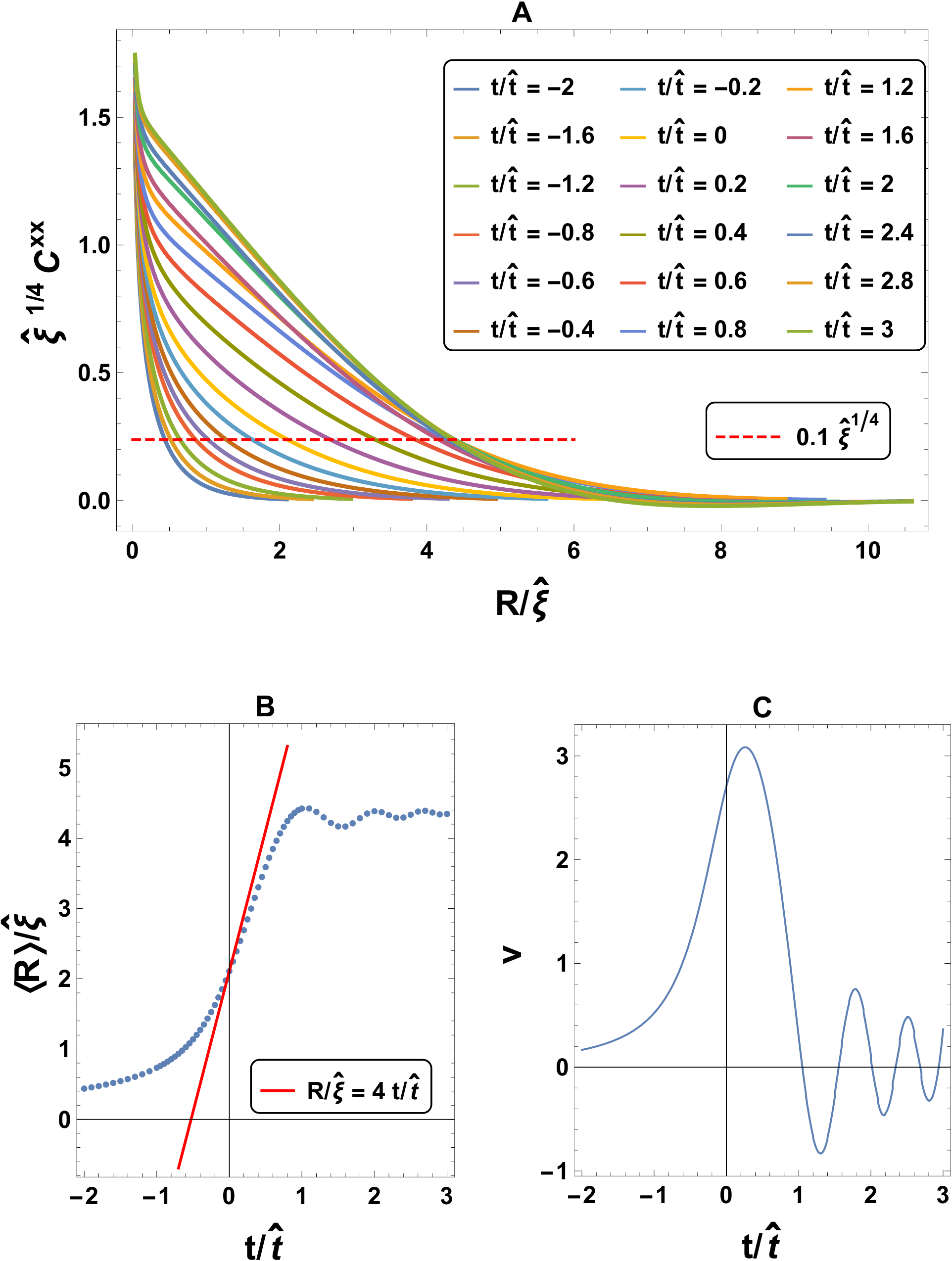}
\caption{  
In A,
the ferromagnetic correlation function in the quantum Ising chain (\ref{Hsigma}),
$C^{xx}_R=\langle \sigma^x_n \sigma^x_{n+R} \rangle-\langle \sigma^x_n\rangle\langle \sigma^x_{n+R}\rangle$, 
as a function of the rescaled distance $R/\hat\xi$ at different rescaled times $t/\hat t=-2,...,3$. 
Here $\sigma^x_n$ is the $x$-component of a spin-$1/2$ at the lattice site $n$
and $\langle \sigma^x_n \rangle$ is the ferromagnetic magnetization (order parameter).
The quench time is $\tau_Q=1024$.
In B,
a correlation range $\langle R \rangle$ as a function of time. We define $\langle R\rangle$ as the distance $R$ where the correlator $C_R^{xx}$ falls below $0.1$. This cutoff is somewhat arbitrary. The behaviour of $\langle R\rangle$ is similar for other cutoffs of this order. It is selected to provide the size of domains that choose the same broken symmetry state (rather than as an estimate of the correlation length, which is defined for much smaller values of $C_R^{xx}$. The range $\langle R\rangle$ increases several times between $-\hat t$ and $\hat t$ before it freezes after $\hat t$.
Similar plots as A and B were made in Ref. \cite{Kolodrubetz}.
In C, 
we show the velocity $v=d\langle R/\hat\xi \rangle/d(t/\hat t)$ --
the derivative of the plot in B -- 
as a function of the rescaled time.
Since in the Ising chain $z=1$ and $\hat\xi=\hat t$,
it is also the velocity in the standard units,
$v=d\langle R \rangle/dt$. 
Near $t/\hat t\approx0$ the correlation is spreading with nearly twice the speed of quasiparticles, $c=2$, 
at the critical point, suggesting that ``sound cones'' (in analogy with ``light cones'') are responsible 
for size of the domains.
}
\label{figCxx}
\end{figure}

\section{ Quasiparticles and sonic horizon} \label{sec:horizon}

However, the reality turns out to be more interesting. 
In the following we will see that all scaling functions do depend on $t/\hat t$ during the non-adiabatic stage. 
For instance, 
in Figure \ref{figCxx} we show the ferromagnetic correlation function in the quantum Ising chain. 
Near $t/\hat t=0$ its range grows almost as the size of the ``sound cone'' -- with twice the speed of quasiparticles at the critical point. 
Between $t/\hat t=-1$ and $t/\hat t=1$ it has enough time to increase several times.
The quench excites entangled pairs of quasiparticles with opposite quasimomenta that spread correlations across the system \cite{quasiparticlehorizon}.
This sonic horizon effect is in a sense at odds with the simple-minded narrative of the impulse approximation. Indeed, as the correlation range grows with time, 
it appears to undermine the significance of $\hat\xi$ as a preferred scale of length. 
Nonetheless,
in the following we will see the KZM scaling holds with $\hat\xi$ as the relevant length.

In order to relate scaling deduced from the ``freeze out'' picture implied by the impulse approximation (where the evolution pauses in the interval $[-\hat t, ~ \hat t]$, and the scale $\hat \xi$ is ``inherited'' from the frozen out pre-transition fluctuations) and the view based on causality and sonic horizon, we focus on a quench-induced evolution in the near-critical regime. 
After $-\hat t$ the state must depart from the adiabatic ground state 
as otherwise its correlation length would diverge at the critical point, since correlations cannot spread infinitely fast. 
Respecting this speed limit, 
after the freezeout at $ -\hat t$ the range of correlations continues to grow, 
but with a finite speed set by
\be
\hat v\simeq\hat\xi/\hat t
\ee 
given by a combination of the relevant scales that defines the speed of the relevant sound. 
Indeed, 
the non-adiabatic evolution excites low-frequency quasiparticles with quasimomenta up to 
\be 
\hat k \simeq \hat\xi^{-1}.
\ee
For a quasiparticle dispersion $\propto k^z$ at the critical point, 
the maximal velocity of the excitations is
\be 
\hat v \sim 
\hat k^{z-1} =
\frac{\hat k^z}{\hat k} \simeq 
\frac{\hat\xi}{\hat t} \sim
\tau_Q^{-(z-1)\nu/(1+z\nu)}.
\ee
With twice this velocity, 
the correlation length can grow from the initial $\hat\xi$ near $-\hat t$ to a final
$
\hat \xi +(2\hat t)(2\hat v)=5\hat\xi
$
near $\hat t$. The final length, even though multiplied by factor of $\sim5$, it still proportional to the original $\hat\xi$.

A few remarks are in order before we begin to illustrate this discussion with the example of the Ising chain. 
We first note that even though the impulse approximation is not accurate in general,
occasionally it yields remarkably accurate, or even exact, results \cite{DamskiImpulse}.
The correlation range of $\sim 5\hat\xi$ may help explain some of the discrepancy between simple estimates of defect density and numerical simulations 
(where it was noted that defects are separated by distances of several $\hat \xi$ (see e.g. \cite{LagunaZ1,YZ, ABZ99}). 
Last but not least, 
we also note that the behavior of the speed of sound in the near-critical regime is controlled by the dynamical critical exponent $z$. 
In the quantum Ising chain $z=1$, which means that the speed of sound is constant with respect to the quench time.
We can however envisage situations where propagation of quasiparticles is impeded (e.g., by damping or conservation laws).
That would complicate the sonic horizon scenario, and could even make the ``freeze-out paradigm'' an accurate approximation.

\section{ Quantum Ising chain } \label{sec:Ising}

We test the KZM scaling in the quantum Ising chain
\be
H~=~-\sum_{n=1}^N \left( g\sigma^z_n + \sigma^x_n \sigma^x_{n+1} \right)~
\label{Hsigma}
\ee
with periodic boundary conditions. For $N\to\infty$ it has two critical points at
$g_c=\pm 1$ between a ferromagnetic phase when $|g|<1$ and two paramagnetic phases when 
$|g|>1$. We assume $g>0$ for definiteness.

A linear quench runs from $t=-\infty$ and across the critical point when $t=0$:
\be
g(t)~=~1-\frac{t}{\tau_Q}~=~1-\epsilon(t)~.
\label{glinear}
\ee
The critical exponents are $z=\nu=1$. The KZM yields the temporal and spatial scales:
\be 
\hat t  \simeq \sqrt{\tau_Q},~~
\hat\xi \simeq \sqrt{\tau_Q}.
\ee
In the following exact solutions,
we will use definitions $\hat t\equiv\sqrt{\tau_Q}$ and $\hat\xi\equiv\sqrt{\tau_Q}$.

\subsection{ From spins to Landau-Zener model} \label{sec:LZ}

Here we assume that $N$ is even for convenience. Following the Jordan-Wigner transformation,
\bea
&&
\sigma^x_n~=~
 -
 \left( c_n + c_n^\dagger \right)
 \prod_{m<n}(1-2 c^\dagger_m c_m)~, \\
&&
\sigma^y_n~=~
 i
 \left( c_n - c_n^\dagger \right)
 \prod_{m<n}(1-2 c^\dagger_m c_m)~, \\
&&
\sigma^z_n~=~1~-~2 c^\dagger_n  c_n~, 
\label{JordanWigner}
\eea
we introduce fermionic operators $c_n$ that satisfy
$\left\{c_m,c_n^\dagger\right\}=\delta_{mn}$ and 
$\left\{ c_m, c_n \right\}=\left\{c_m^\dagger,c_n^\dagger \right\}=0$. 
The Hamiltonian (\ref{Hsigma}) becomes
\be
 H~=~P^+~H^+~P^+~+~P^-~H^-~P^-~.
\label{Hc}
\ee
Above
$
P^{\pm}=
\frac12\left[1\pm P\right]
$
are projectors on subspaces with even ($+$) and odd ($-$) parity
\be 
P~=~
\prod_{n=1}^N\sigma^z_n ~=~
\prod_{n=1}^N\left(1-2c_n^\dagger c_n\right) ~
\ee
and  
\bea
H^{\pm}=
\sum_{n=1}^N
\left( 
g c_n^\dagger  c_n - c_n^\dagger  c_{n+1} - c_{n+1}  c_n - \frac{g}{2}
\right)
+{\rm H.c.}
\label{Hpm}
\eea
are corresponding reduced Hamiltonians. The $c_n$'s in $H^-$ satisfy
periodic boundary condition $c_{N+1}=c_1$, but the $c_n$'s in $H^+$
are anti-periodic: $c_{N+1}=-c_1$.

The initial ground state at $g\to\infty$ has even parity, hence we can focus on the even subspace. $H^+$ is diagonalized by a Fourier transform followed by a Bogoliubov transformation. The anti-periodic Fourier transform is  
\be
c_n~=~ 
\frac{e^{-i\pi/4}}{\sqrt{N}}
\sum_k c_k e^{ikn}~,
\label{Fourier}
\ee
where the pseudomomentum takes half-integer values
\be
k~=~
\pm \frac12 \frac{2\pi}{N},
\dots,
\pm \frac{N-1}{2} \frac{2\pi}{N}~.
\label{halfinteger}
\ee
The Hamiltonian (\ref{Hpm}) becomes
\bea
&&
H^+~ ~=~
\nonumber\\
&&
\sum_k
\left[
2(g-\cos k) c_k^\dagger c_k +
\sin k
\left( 
 c^\dagger_k c^\dagger_{-k}+
 c_{-k} c_k
\right)-
g
\right]~.
\label{Hck}
\eea
Its diagonalization is completed by a Bogoliubov transformation
$
c_k=U_k\gamma_k+V_{-k}^*\gamma^\dagger_{-k},
$
provided that Bogoliubov modes $(U_k,V_k)$ are eigenstates of the stationary
Bogoliubov-de Gennes equations
\bea
\omega_k 
\left(
\begin{array}{c}
U_k \\
V_k
\end{array}
\right) &=&
2\left[\sigma^z(g-\cos k)+\sigma^x\sin k\right]
\left(
\begin{array}{c}
U_k \\
V_k
\end{array}
\right)~. 
\label{stBdG}
\eea
with a positive eigenfrequency
\be
\omega_k~=~2\sqrt{(g-\cos k)^2+\sin^2 k}~.
\label{epsilonkIsing}
\ee
The corresponding normalized eigenstate $\left(U_k,V_k\right)$ defines a quasiparticle operator, 
$
\gamma_k=U_k c_k + V_{-k} c_{-k}^\dagger, 
$
bringing the Hamiltonian to the diagonal form
$
H^+=E_0^{+}+\sum_k \omega_k\gamma_k^\dagger \gamma_k.
$
Thanks to the projection $P^+H^+P^+$ in Eq.~(\ref{Hc}) only states 
with even numbers of quasiparticles belong to the spectrum of $H$ -- in a periodic chain kinks must be created in pairs.

The initial ground state at $g\to\infty$ is a Bogoliubov vacuum $|0\rangle$ annihilated by all $\gamma_k$. 
As $g(t)$ is ramped down, the state gets excited from the instantaneous ground state, but in the Heisenberg 
picture it remains the initial vacuum. Instead, the fermionic operators are time-dependent
\be
c_k ~=~ u_k(t) \gamma_k + v_{-k}^*(t) \gamma^\dagger_{-k}~,
\label{tildeBog}
\ee
with the initial condition $(u_k,v_k)=(1,0)$. They satisfy Heisenberg equations
$
i\frac{d}{dt} c_k~=~\left[ c_k, H^+\right]
$
equivalent to the time-dependent Bogoliubov-de Gennes equations (\ref{stBdG}):
\bea
i\frac{d}{dt} 
\left(
\begin{array}{c}
u_k \\
v_k
\end{array}
\right) &=&
2\left[\sigma^z[g(t)-\cos k]+\sigma^x\sin k\right]
\left(
\begin{array}{c}
u_k \\
v_k
\end{array}
\right). 
\label{dynBdG}
\eea
A new time variable for $k>0$,
\be
t'~=~4\tau_Q\sin k\left(-1+\frac{t}{\tau_Q}+\cos k\right),
\ee
brings Eqs. (\ref{dynBdG}) to the canonical LZ form
\bea
i\frac{d}{dt'}
\left(
\begin{array}{c}
u_k \\
v_k
\end{array}
\right) &=&
\frac12
\left[-\frac{t'}{\tau_k}\sigma^z+\sigma^x\right]
\left(
\begin{array}{c}
u_k \\
v_k
\end{array}
\right)~,
\label{BdGLZ}
\eea
with a transition time $\tau_k=4\tau_Q\sin^2 k$. The solution of Eqs. (\ref{dynBdG}) is
\bea
u_k &=&
e^{-\frac{\pi}{16}\tau_k}
D_{\frac14i\tau_k}(z)
e^{i\pi/4},\\
v_k &=&
\frac12
e^{-\frac{\pi}{16}\tau_k}
D_{-1+\frac14i\tau_k}(z)
\sqrt{\tau_k}.
\label{general} 
\eea
Here $D_m(z)$ is the Weber function with an argument
\be 
z=e^{3\pi i/4}\frac{t'}{\sqrt{\tau_k}}.
\ee
The scaling is not apparent in this exact formula.
 
\begin{figure}[t]
\includegraphics[width=1.0\columnwidth,clip=true]{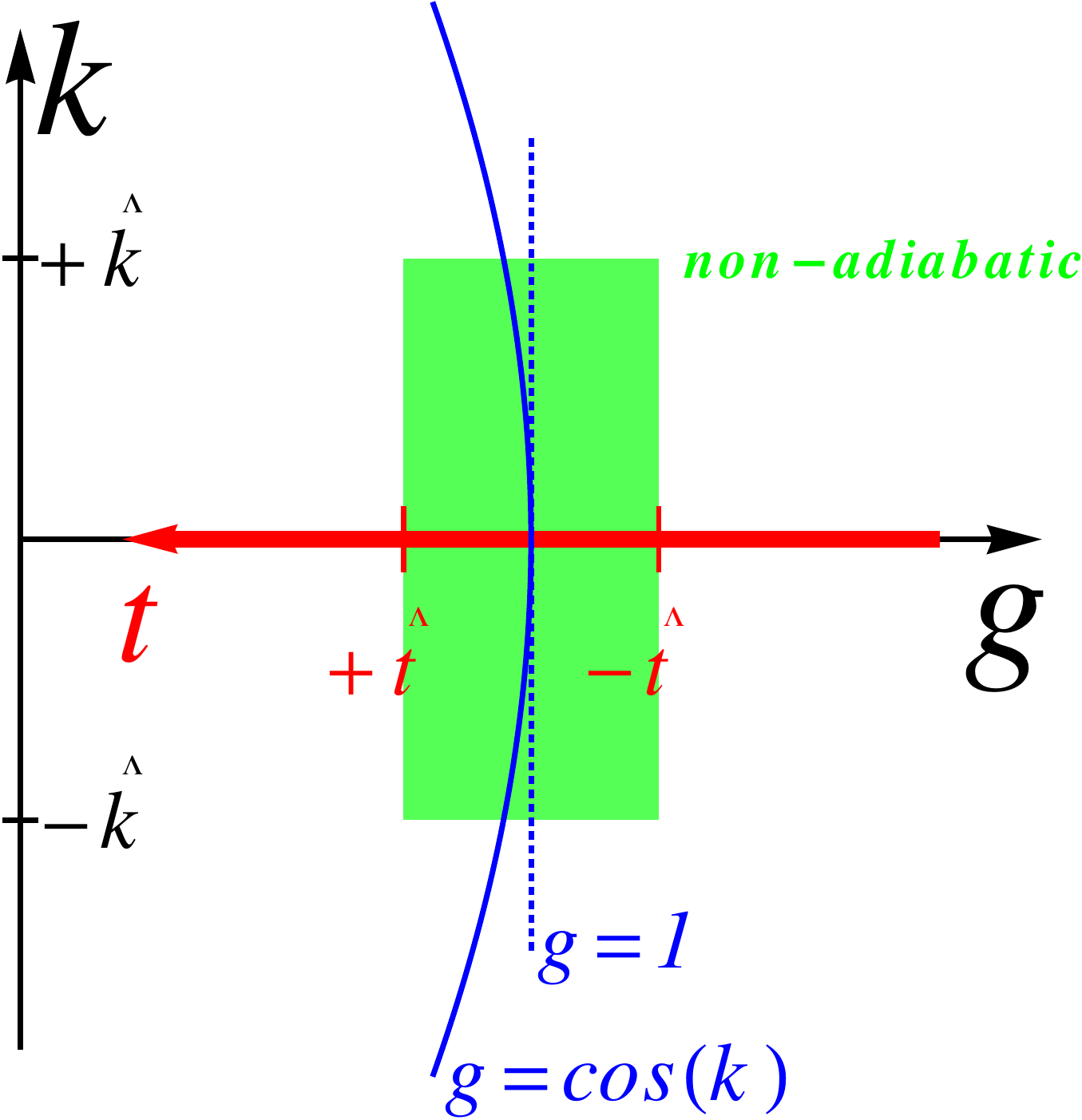}
\caption{  
The Landau-Zener modes (\ref{dynBdG}) with different $k$ pass through their anti-crossings at different 
$g_{\rm AC}(k)=\cos(k)$.
In the adiabatic limit, 
the size of the non-adiabatic regime $\hat k\simeq 1/\sqrt{\tau_Q}$ becomes small,
$1-g_{\rm AC}(\hat k)\simeq1/\tau_Q$ becomes much less than $1-g(\hat t)\simeq1/\sqrt{\tau_Q}$,
and we can safely approximate $g_{\rm AC}(k)\approx1$, 
as if all the anti-crossings took place simultaneously at the critical $g=1$.
This is the essence of the approximation between Eqs. (\ref{rescuv}) and  (\ref{infuv}). 
}
\label{LZfigure}
\end{figure}
 
\subsection{ Scaling in Landau-Zener model }\label{sec:LZscaling}

Only small quasimomenta up to 
\be
\hat k = 1/\sqrt{\tau_Q}
\ee 
get excited. For $k\ll1$ we can approximate
\bea
u_k &=&
e^{-\frac14\pi q^2}
D_{iq^2}(z)
~e^{i\pi/4},\nonumber\\
v_k &=&
e^{-\frac14\pi q^2}
D_{-1+iq^2}(z)
~q,\nonumber\\
z &=&
2e^{3\pi i/4}\left(\frac{t}{\hat t}-\frac{q^2}{2\sqrt{\tau_Q}}\right),
\label{rescuv} 
\eea
where
$
q = k/\hat k
$
is a rescaled quasimomentum.
 
Only $q$ up to $q\approx1$ get excited. For them, when $\tau_Q$ is large enough,
we can further approximate $z\approx 2e^{3\pi i/4}(t/\hat t)$, see Figure \ref{LZfigure},
and obtain
\bea
u_k &=&
e^{-\frac14\pi q^2}
D_{iq^2}\left(2e^{3\pi i/4}\frac{t}{\hat t}\right)
~e^{i\pi/4},\nonumber\\
v_k &=&
e^{-\frac14\pi q^2}
D_{-1+iq^2}\left(2e^{3\pi i/4}\frac{t}{\hat t}\right)
~q.
\label{infuv} 
\eea
As required by the space-time scaling, 
these non-adiabatic modes depend on the rescaled $t/\hat t$ and $q$ only.

\begin{figure*}[t]
\includegraphics[width=1.95\columnwidth,clip=true]{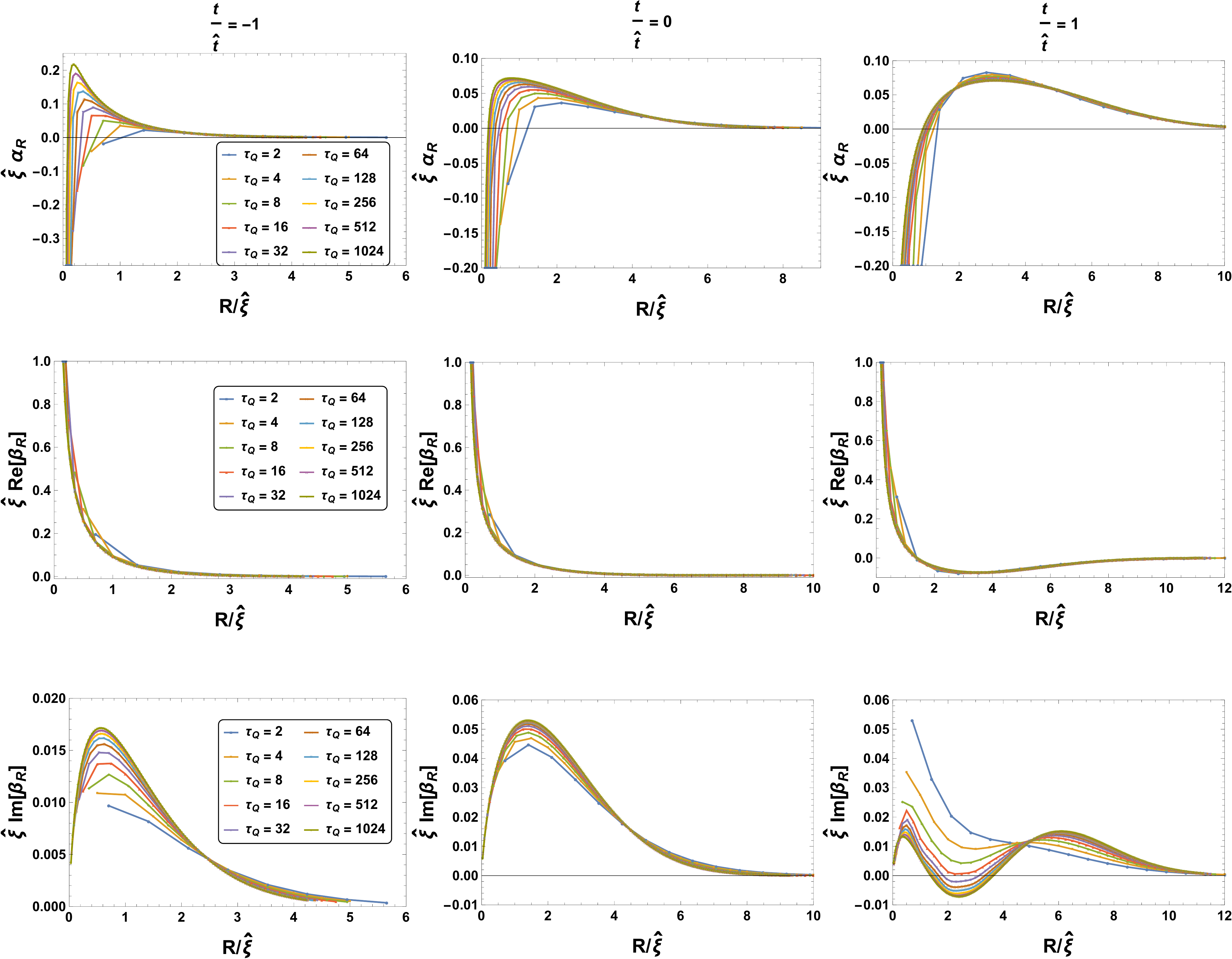}
\caption{  
Quadratic fermionic correlators in Eqs. (\ref{alphaR}) and (\ref{betaR}).
The first row shows the real $\alpha_R=\langle c_R c_0^\dag \rangle$. 
The second and third rows show the real and imaginary parts of 
$\beta_R=\langle c_R c_0 \rangle$, respectively.
The left, middle and right columns show the correlators before ($t/\hat t=-1$),
at ($t/\hat t=0$), and after ($t/\hat t=+1$) the critical point, respectively. 
Different colors of the plots correspond to different quench times $\tau_Q$.
All plots are rescaled: they are in function of the rescaled distance $R/\hat\xi=R/\sqrt{\tau_Q}$ and the correlators are multiplied with 
$\hat\xi=\sqrt{\tau_Q}$.
At large distance, $R\gg1$,
the plots in all nine panels collapse asymptotically with increasing $\tau_Q$ demonstrating the scaling hypotheses (\ref{alphascaling}) and (\ref{betascaling}) 
for large enough $\tau_Q$.
The collapse does not happen at short distance, $R\approx 1$, where the microscopic lattice constant remains a relevant scale of length and
the scaling hypothesis is not expected to hold.
The collapsed plots for large $\tau_Q$ are the scaling functions
$F_\alpha\left(t/\hat t,R/\hat\xi\right)$ and $F_\beta\left(t/\hat t,R/\hat\xi\right)$.
}
\label{fig:alphabeta1}
\end{figure*}

\subsection{ Scaling in fermionic correlators  }\label{sec:scalingfermions}

The state during the quench is fully determined by time-dependent quadratic correlators.
In the thermodynamic limit $N\to\infty$,
they are given by integrals:
\bea
\alpha_{R}(t) &\equiv& 
   \langle c_R c_0^\dagger \rangle ~=~ 
   \frac{1}{\pi}\int_{0}^\pi dk~|u_k|^2~\cos(k R)~,
\label{alphaR}\\
\beta_{R}(t) &\equiv& 
   \langle c_R c_0 \rangle ~=~ 
   \frac{1}{\pi}\int_{0}^\pi dk~u_kv_k^*~\sin(k R).
\label{betaR}
\eea
The integrals extend into the adiabatic regime, where the scaling form (\ref{infuv}) is no longer applicable. 
Instead, the modes $u_k,v_k$ can be approximated (up to an irrelevant dynamical phase) by the adiabatic eigenmodes 
$U_k,V_k$ at $g=1-(t/\hat t)/\sqrt{\tau_Q}$.

In order to demonstrate the scaling of $\alpha_R$, it is convenient to rearrange it first as
\bea
\alpha_{R} &=& \alpha_R^{\rm (KZ)} + \alpha_R^{\rm (GS)} + \alpha_R^{\rm (cr)} , \label{alpha}
\eea
where
\bea
\alpha_R^{\rm (KZ)} &=& \frac{1}{\pi}\int_{0}^\pi dk ~ \left( |u_k|^2 - |U_k|^2 \right)\cos(k R),            \label{alphaKZ}\\
\alpha_R^{\rm (GS)} &=& \frac{1}{\pi}\int_{0}^\pi dk ~ \left( |U_k|^2 - |U_k^{\rm CP}|^2 \right)\cos(k R),    \label{alphaGS}\\
\alpha_R^{\rm (CP)} &=& \frac{1}{\pi}\int_{0}^\pi dk ~ |U_k^{\rm CP}|^2\cos(k R).                          \label{alphacr}
\eea
Here $U_k$ and $U_k^{\rm CP}$ are the adiabatic eigenmodes at $g=1-(t/\hat t)/\sqrt{\tau_Q}$ and the critical $g=1$, respectively.

Since the correlation length in the ground state at $g=1-(t/\hat t)/\sqrt{\tau_Q}$ is $\xi\simeq\sqrt{\tau_Q}/(t/\hat t)$, then in Eq. (\ref{alphaGS}) the integrand is nonzero up to $\hat k \simeq\xi^{-1}$. Consequently, given that $\xi\simeq\hat\xi/(t/\hat t)$, 
a change of the integration variable $k\to k\hat\xi$ is enough to show that
\be 
\alpha_R^{\rm (GS)}~=~\hat\xi^{-1}~F_\alpha^{\rm (GS)}\left(t/\hat t,R/\hat\xi\right).
\label{alphaGSscaling}
\ee 
Here $F_\alpha$ is a scaling function.

In a similar way,
in Eq. (\ref{alphaKZ}) the integrand is non-zero in the non-adiabatic regime up to $\hat k$.
In this regime $u_k$ has the scaling form (\ref{infuv}) and $U_k$ has a characteristic quasimomentum scale 
$\simeq\xi^{-1}\sim\hat\xi^{-1}$. Consequently,
the same change of the integration variable shows again that
\be 
\alpha_R^{\rm (KZ)} ~=~ \hat\xi^{-1}~F_\alpha^{\rm (KZ)}\left(t/\hat t,R/\hat\xi\right) 
\label{alphaKZscaling}
\ee 
for large enough $\tau_Q$. 

Finally, Eq. (\ref{alphacr}) is the ground-state correlator at the critical point:
\be 
\alpha_R^{\rm (CP)}=
\frac{-1}{2R^2-\frac12}=
\hat\xi^{-2}\frac{-1}{2\left(\frac{R}{\hat\xi}\right)^2-\frac{1}{2\hat\xi^2}}\approx
\hat\xi^{-2}\frac{-\frac12}{\left(R/\hat\xi\right)^2}.
\label{alphacrscaling}
\ee
It has a scaling form, 
but its scaling dimension $-2$ is twice the $-1$ in Eqs. (\ref{alphaGSscaling},\ref{alphaKZscaling}). 
For slow enough quenches $\alpha_R^{\rm (CP)}$ becomes negligible as compared to the other two terms.

Collecting together Eqs. (\ref{alphaGSscaling},\ref{alphaKZscaling},\ref{alphacrscaling}) and (\ref{alpha})
we can conclude with a dynamical scaling law
\be 
\alpha_R ~=~ \hat\xi^{-1}~F_\alpha\left(t/\hat t,R/\hat\xi\right)
\label{alphascaling}
\ee 
valid for large enough $\tau_Q$. In Figure \ref{fig:alphabeta1} we show rescaled plots supporting this conclusion
for large distances $R\gg1$ and the quench time $\tau_Q$ where the scaling hypothesis is expected to hold. 
The plots were obtained by numerical integration in Eq. (\ref{alphaR}), see Appendix \ref{numalpha}.

The argument for $\beta_R$ is similar except that in the critical ground-state the scaling dimension is $-1$:
\be 
\beta_R^{\rm (CP)}=
\frac{1}{4R-\frac{1}{R}}\approx
\hat\xi^{-1}\frac{\frac14}{\left(R/\hat\xi\right)}.
\label{betacrscaling}
\ee
This difference does not alter the overall scaling 
\be 
\beta_R ~=~ \hat\xi^{-1}~F_\beta\left(t/\hat t,R/\hat\xi\right)
\label{betascaling}
\ee 
with the same dimension. Figure \ref{fig:alphabeta1} supports this conclusion for large distances $R\gg1$ and 
the quench time $\tau_Q$ where the scaling hypothesis is expected to hold. 

The quadratic correlators completely determine the Bogoliubov vacuum state. They satisfy the KZM scaling. 
Therefore, it is tantalizing to take the scaling for granted for any operator $O(x)$ in this state.
However, as the quadratic correlators satisfy the scaling only asymptotically for slow enough $\tau_Q$,
we cannot assume that their convergence with $\tau_Q$, or collapse in Fig. \ref{fig:alphabeta1}, 
is fast enough to warrant similar collapse for any operator $O(x)$. Therefore, in the following
we study the most interesting observables case by case.    

\begin{figure}[t]
\includegraphics[width=1\columnwidth,clip=true]{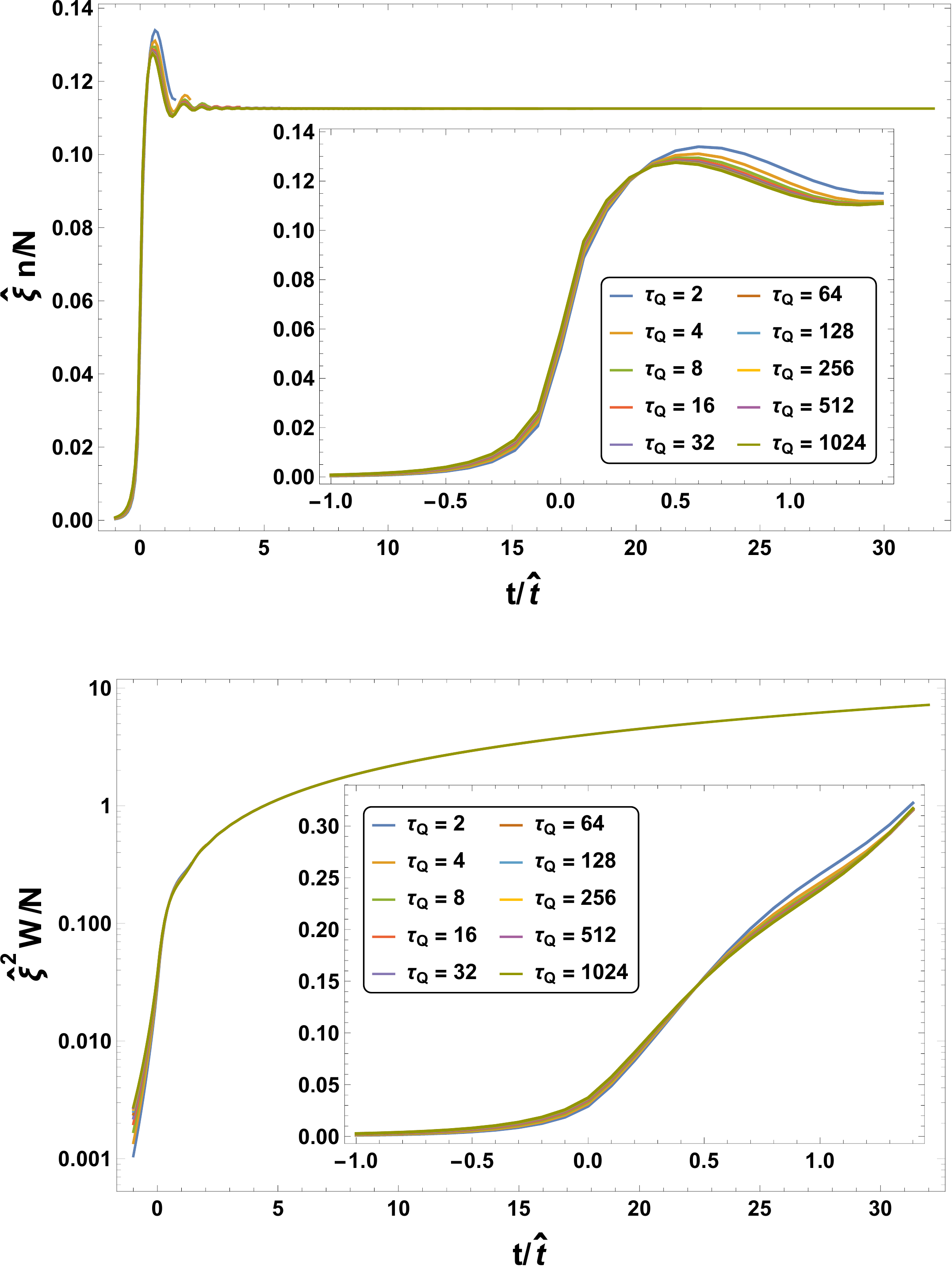}
\caption{ 
Top, 
the density of quasiparticle excitations in Eq. (\ref{nN}) in function of the rescaled time $t/\hat t$. Different colors of the plots correspond to different quench times $\tau_Q$.
Each plot reaches up to $t/\hat t=\sqrt{\tau_Q}$ corresponding to zero transverse field $g=0$. The inset is a focus on the non-adiabatic stage $t/\hat t=-1...1$.
Bottom,
the density of excitation energy in Eq. (\ref{WN}) in function of $t/\hat t$.  
In both panels,
the plots collapse asymptotically with increasing $\tau_Q$ demonstrating the scaling hypotheses (\ref{FnW}) for large enough $\tau_Q$.
The collapsed plots for large $\tau_Q$ are the scaling functions
$F_n\left(t/\hat t\right)$ and $F_W\left(t/\hat t\right)$ in Eq. (\ref{FnW}).
Same quantities with a different re-scaling of time can be seen in Fig. \ref{gscaling}.
}
\label{nWrescaled}
\end{figure}
\subsection{ Scaling in energy and number of excitations }\label{sec:scalingenergy}

To begin with operators that do not depend on any distance $x$, we consider
density of quasiparticle excitations,
\be 
\frac{n}{N} = \frac{1}{\pi} \int_0^{\pi} dk~ p_k,
\label{nN}
\ee
and excitation energy,
\be 
\frac{W}{N} = \frac{1}{\pi} \int_0^{\pi} dk~ p_k ~2\omega_k,
\label{WN}
\ee
both in the thermodynamic limit $N\to\infty$. Here $\omega_k$ is the instantaneous quasiparticle dispersion 
(\ref{epsilonkIsing}), and $p_k$ is excitation probability for a pair of quasiparticles with quasimomenta
$(k,-k)$:
\be 
p_k=\left| 
(-V_k,U_k)\left(\begin{array}{c}u_k\\ v_k\end{array}\right)
\right|^2.
\ee
Since $p_k$ is non-zero in the non-adiabatic regime only up to $\hat k$ and, furthermore, $\omega_k\sim k$ 
in this regime, a change of the integration variable from $k$ to $k\hat\xi$ leads to the scaling forms:
\be 
\hat\xi^1 \frac{n}{N}=F_n(t/\hat t), ~~
\hat\xi^2 \frac{W}{N}=F_W(t/\hat t).
\label{FnW}
\ee 
The last form is consistent with the prediction of Ref. \cite{Polkovnikov11} for gapless
systems. The collapsing plots in Fig. \ref{nWrescaled}, similar to the plots in Ref. \cite{ViolaOrtiz}, demonstrate this scaling. Interestingly, the work density collapses well beyond $t/\hat t=1$ even though the gapless $\omega_k\sim k$ does not apply there.

In order to understand why, notice that the excitation probability is a scaling function 
$p_k(t)=p(t/\hat t,k/\hat k)$ that is non-zero only up to $k\approx\hat k\equiv1/\sqrt{\tau_Q}$.
In this regime of small $k$ the dispersion (\ref{epsilonkIsing}) is 
$\omega_k\approx2\sqrt{\epsilon^2+k^2-\epsilon k^2}$, where $\epsilon=1-g=t/\tau_Q$. 
With a new integration variable $q=k/\hat k$ Eq. (\ref{WN}) becomes
\be 
\frac{W}{N} = 
      \frac{2\hat k^2}{\pi} \int_0^\infty dq~ 
      p\left(\frac{t}{\hat t},q\right)
      \sqrt{\left(\frac{t}{\hat t}\right)^2+q^2-\frac{q^2}{\sqrt{\tau_Q}}\left(\frac{t}{\hat t}\right)}
\ee
In the adiabatic limit the last term under the square root becomes negligible and the right 
hand side becomes $\hat\xi^{-2} F_W(t/\hat t)$, i.e., a scaling function of $t/\hat t$ only.
For a given $t/\hat t$, the excitation energy scales like $W/N\sim\hat\xi^{-2}=\tau_Q^{-1}$.

\begin{figure}[t]
\includegraphics[width=.99\columnwidth,clip=true]{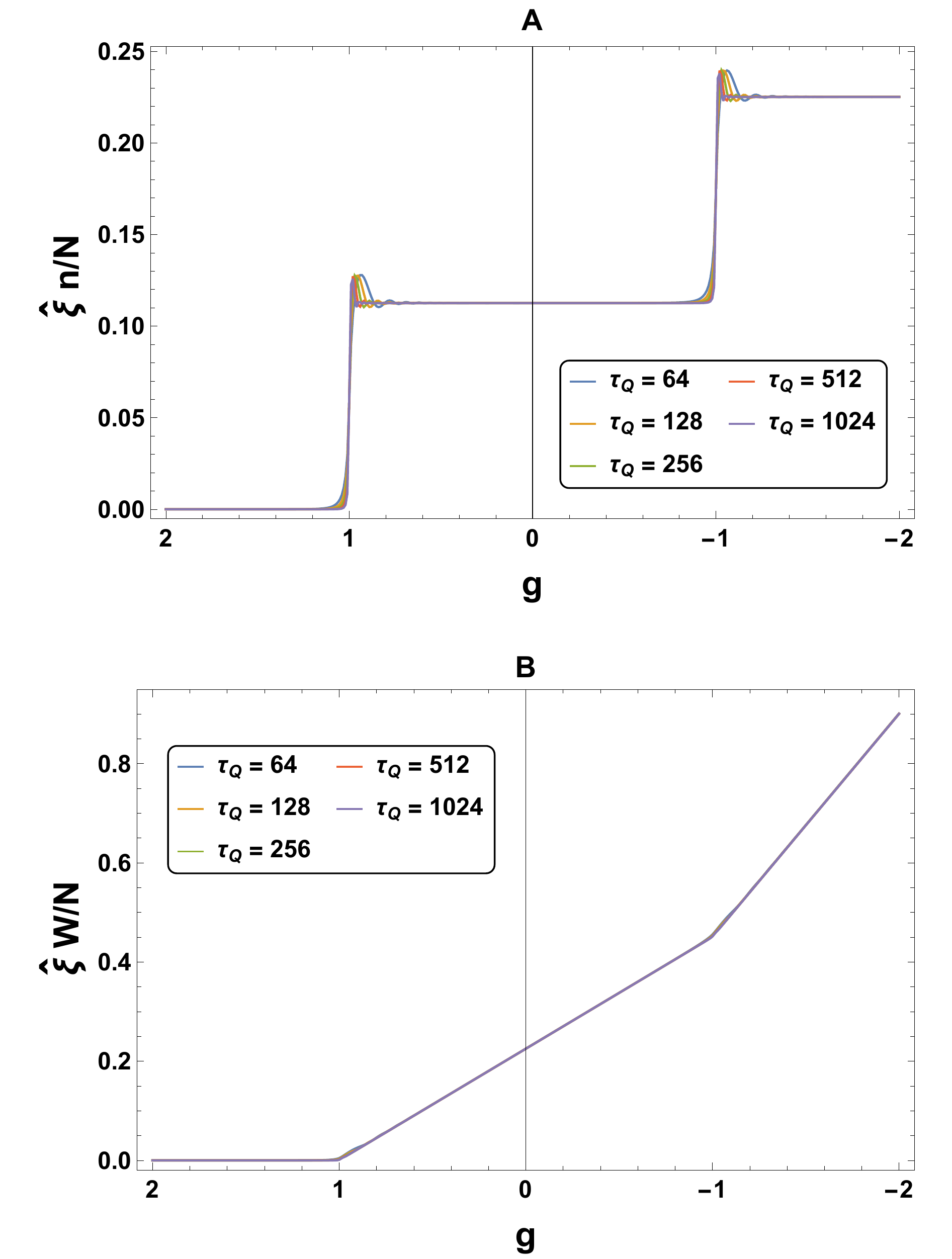}
\caption{  
In A,
the density of quasiparticle excitations as a function of $g$.
Quasiparticles are excited at the two critical points, $g=\pm1$.
For any fixed $g$, 
except near $g=\pm1$,
their density scales like $n/N\sim\hat\xi^{-1}$ (i.e., the typical distance between the kinks is set by $\hat\xi$).
In B,
the density of excitation energy as a function of $g$.
For any fixed $g$ it scales like $W/N\sim\hat\xi^{-1}$,
except near the gapless critical points $g=\pm1$. Apart from the immediate vicinity of the critical points the evolution is adiabatic. That is, the increase of energy is caused by the increase in the separations between the occupied energy levels. Thus, when Ising chain is still in its ground state (leftmost segment in B), $W=0$, but after the excitation caused by the passage through the critical point at $g=1$, the slope increases, and remains constant till $g=-1$. Additional excitations double the slope in the rightmost segments of B. Indeed, these slopes are set by the quasiparticle densities seen above in A.
}
\label{gscaling}
\end{figure}

This seems to contradict Ref. \cite{Dziarmaga2005} where the excitation energy at $g=0$ (proportional 
to the number of kinks) scales like $W/N\sim\hat\xi^{-1}=\tau_Q^{-1/2}$. However, there is no contradiction, 
since the two scalings compare energies for different $\tau_Q$ either at a constant $t/\hat t$ or a constant $g$. 
At the constant $g=0$, corresponding to the $\tau_Q$-dependent $t/\hat t=\sqrt{\tau_Q}$, we have a flat dispersion 
$\omega_k=2$ in Eq. (\ref{WN}) and the excitation energy is proportional to the number of quasiparticle excitations 
$W/N=4n/N\sim\hat\xi^{-1}=\tau_Q^{-1/2}$. For illustration, in Figure \ref{gscaling} we show the quasiparticle and 
energy densities as a function of $g$ instead of $t/\hat t$. Everywhere except near the gapless critical points, 
for a fixed $g$ the energy scales like $W/N\sim\hat\xi^{-1}=\tau_Q^{-1/2}$. This is a remarkable change
of perspective, even though the picture away from criticality is sensitive to relevant/non-integrable
perturbations of the Ising model.

\begin{figure*}[t]
\includegraphics[width=1.80\columnwidth,clip=true]{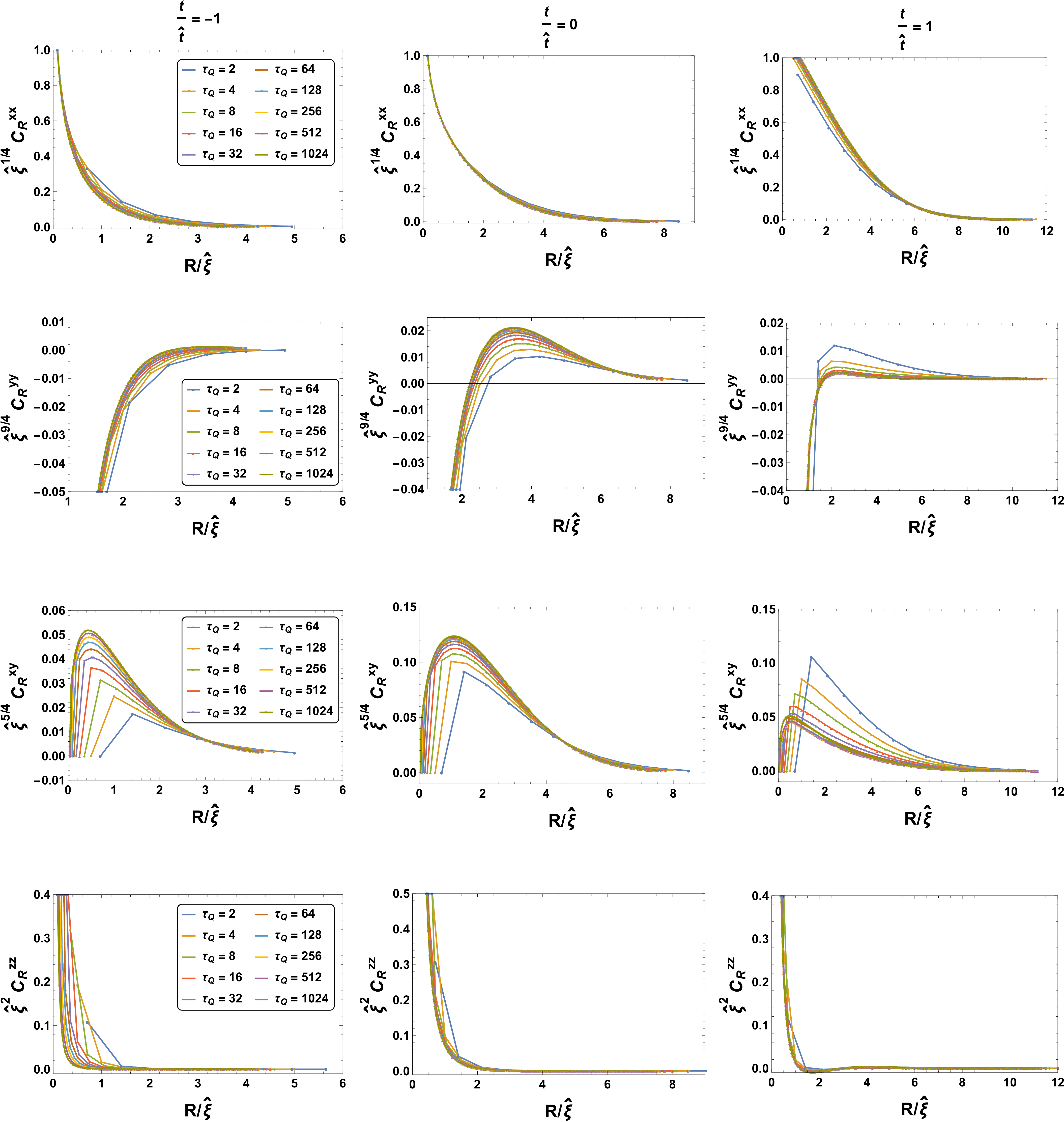}
\caption{  
Spin-spin correlation functions in Eq. (\ref{Cab}).
The left, middle and right columns show the correlators before ($t/\hat t=-1$), at 
($t/\hat t=0$), and after ($t/\hat t=+1$) the critical point, respectively. 
The first row shows the strongest ferromagnetic correlator $C^{xx}_R$,
the second and third one show $C^{yy}_R$ and $C^{xy}_R$, respectively,
and the bottom one shows the transverse $C^{zz}_R$.
Different colors of the plots correspond to different quench times $\tau_Q$.
All plots are rescaled: they are in function of the rescaled distance $R/\hat\xi=R/\sqrt{\tau_Q}$ and the correlators are multiplied with 
$\hat\xi^{\Delta_a+\Delta_b}=\left(\sqrt{\tau_Q}\right)^{\Delta_a+\Delta_b}$.
At large distance,
$R\gg1$,
the plots in all twelve panels collapse asymptotically with increasing $\tau_Q$ demonstrating the space-time scaling (\ref{Cab}) for large enough $\tau_Q$.
The collapse does not happen at short distance, $R\approx 1$, where the microscopic lattice constant remains a relevant scale of length and
the scaling hypothesis is not expected to hold.
The collapsed plots for large $\tau_Q$ are the scaling functions
$F_C^{ab}\left(t/\hat t,R/\hat\xi\right)$ in Eq. (\ref{Cabscaling}).
}
\label{CRrescaled}
\end{figure*}
\subsection{ Scaling in two-spin correlators }\label{sec:scalingCR}

The quadratic fermionic correlators are the building blocks for spin correlators:
\bea 
C^{ab}_R(t) \equiv
\langle \sigma_n^a \sigma_{n+R}^b \rangle-
\langle \sigma_n^a \rangle
\langle \sigma_{n+R}^b \rangle .
\label{Cab}        
\eea 
Except for the transverse $C^{zz}$, they are Pfaffians of matrices whose elements are 
the fermionic correlators (\ref{alphaR},\ref{betaR}), see Ref. \cite{BarouchMcCoy}.

The KZ scaling implies that
\bea 
C^{ab}_R(t) =   
\hat\xi^{-\Delta_a-\Delta_b} ~
F_C^{ab}\left(t/\hat t, R/\hat\xi\right). 
\label{Cabscaling}        
\eea 
Here $\Delta_a$ is the scaling dimension for the operator $\sigma^a$.
In the Ising chain we have $\Delta_x=\frac18$, $\Delta_y=\frac98$, and $\Delta_z=1$.
Figure \ref{CRrescaled} shows rescaled plots of all non-zero correlators. 
Their collapse for large enough $\tau_Q$ confirms the space-time scaling for large distances $R\gg1$ 
where the scaling hypothesis is expected to hold. 

\subsection{ Scaling in mutual information }\label{sec:scalingmutual}

\begin{figure*}[t]
\includegraphics[width=1.95\columnwidth,clip=true]{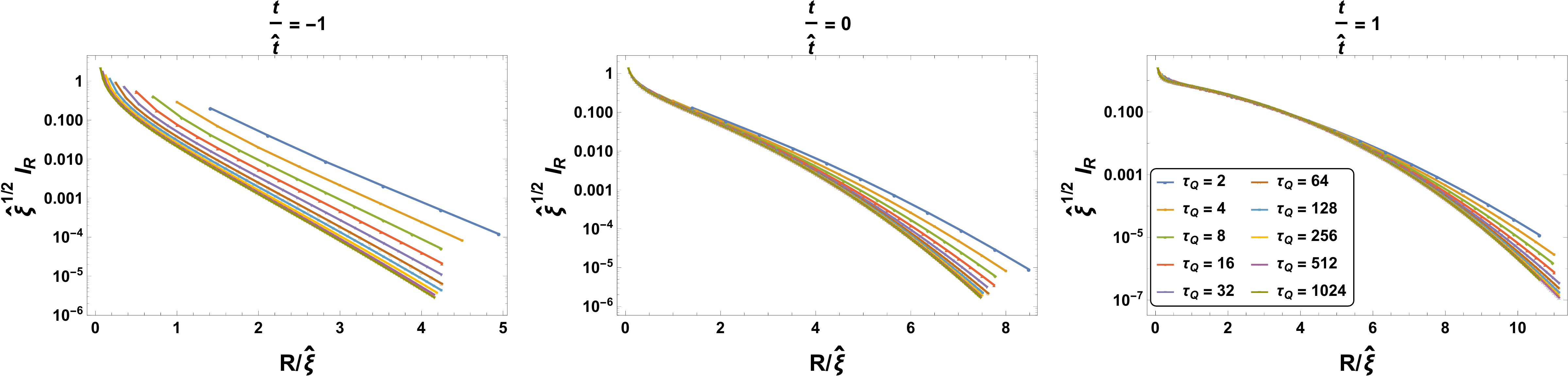}
\caption{  
Mutual information in Eq. (\ref{IR}).
The left, middle and right panels show the mutual information before ($t/\hat t=-1$), at 
($t/\hat t=0$), and after ($t/\hat t=+1$) the critical point, respectively. 
Different colors of the plots correspond to different quench times $\tau_Q$.
All plots are rescaled: 
they are in function of the rescaled distance $R/\hat\xi=R/\sqrt{\tau_Q}$ and 
the mutual information is multiplied with 
$\hat\xi^{4\Delta_a}=\left(\sqrt{\tau_Q}\right)^{1/2}$.
The plots in all three panels collapse asymptotically with increasing $\tau_Q$ demonstrating the space-time scaling (\ref{IRscaling}) for large enough $\tau_Q$.
The collapsed plots for large $\tau_Q$ are the scaling function
$F_I\left(t/\hat t,R/\hat\xi\right)$ in Eq. (\ref{IRscaling}).
}
\label{figmutual}
\end{figure*}
\begin{figure*}[t]
\includegraphics[width=1.95\columnwidth,clip=true]{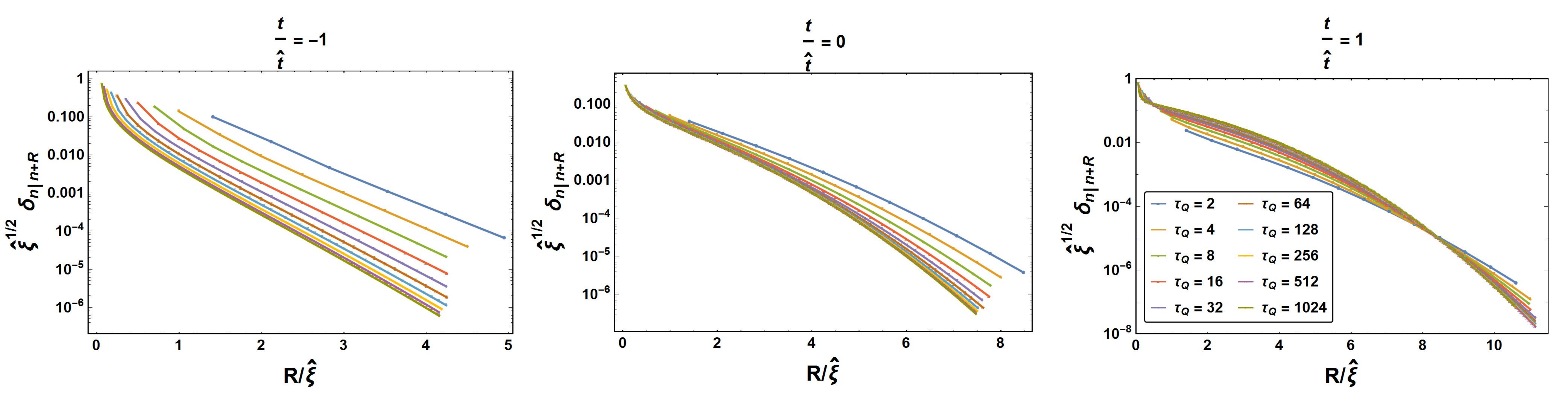}
\caption{  
Quantum discord in Eq. (\ref{discord}).
The left, middle and right panels show the discord before ($t/\hat t=-1$), at 
($t/\hat t=0$), and after ($t/\hat t=+1$) the critical point, respectively. 
Different colors of the plots correspond to different quench times $\tau_Q$.
All plots are rescaled: 
they are in function of the rescaled distance $R/\hat\xi=R/\sqrt{\tau_Q}$ and 
the discord is multiplied with $\hat\xi^{4\Delta_a}=\left(\sqrt{\tau_Q}\right)^{1/2}$.
The plots in all three panels collapse asymptotically with increasing $\tau_Q$ demonstrating the space-time scaling (\ref{discordscaling}) for large enough $\tau_Q$.
The collapsed plots for large $\tau_Q$ are the scaling function
$F_{\delta}\left(t/\hat t,R/\hat\xi\right)$ in Eq. (\ref{discordscaling}).
}
\label{figdiscord}
\end{figure*}

The overall strength of spin-spin correlations can be conveniently characterized by mutual information between the two spins. 
A reduced density matrix for the $n$-th spin is
\bea
\rho^{(1)}_n &=& 
\frac12
\left(
1_n+\langle\sigma^z\rangle~ \sigma^z_n
\right).
\eea
A reduced density matrix for spins $n$ and $n+R$ includes their correlations:
\bea 
\rho^{(2)}_{n,n+R} &=&
\rho^{(1)}_n\otimes\rho^{(1)}_{n+R} +
\frac14
\sum_{a,b=1}^3 C_R^{ab}~\sigma^a_n\otimes\sigma^b_{n+R}.
\eea
The correlations contribute to non-zero mutual information between the spins,
\be 
I_R=
S\left[\rho^{(1)}_n\right]+
S\left[\rho^{(1)}_{n+R}\right]-
S\left[\rho^{(2)}_{n,n+R}\right].
\label{IR}
\ee
Here $S[\rho]=-{\rm Tr}\rho\log\rho$ is the von Neumann entropy.

When the correlations $C^{ab}_R$ are weak, for large $R$ or large $\tau_Q$ or both, then they are a small perturbation to 
the uncorrelated product $\rho^{(1)}_n\otimes\rho^{(1)}_{n+R}$. To leading order, the mutual information is
a quadratic form in $C_R$'s whose coefficients depend on the transverse magnetization $\langle\sigma^z\rangle$. 
For slow enough $\tau_Q$, 
the magnetization can be approximated by its value in the ground state at the critical point, 
$\langle\sigma^z\rangle\approx 2/\pi$, and it is enough to keep only the dominant term that is quadratic in the strongest correlator $C^{xx}_R$:
\bea 
I_R &\approx & 
\frac{\pi}{8} \left[ \frac{2 \pi}{\pi^2-4} + {\rm arctanh}\left(\frac{2}{\pi}\right) \right] \left(C^{xx}_R\right)^2 \nonumber\\
&=&
0.72~\left(C^{xx}_R\right)^2.
\eea
Consequently, the mutual information should scale as
\be 
I_R(t)=\hat\xi^{-4\Delta_x}F_I\left(t/\hat t,R/\hat\xi\right),
\label{IRscaling}
\ee
where $F_I\propto \left(F_C^{xx}\right)^2$ is a scaling function.
This scaling is demonstrated by the collapsing plots in Fig. \ref{figmutual}.

\subsection{ Scaling in quantum discord }\label{sec:scalingdiscord}

A convenient measure of quantumness of correlations between spins $n$ and $n+R$ is the quantum discord \cite{discord}:
\be 
\delta_{n|n+R}=
{\rm Min}_\sigma ~S\left[n|\sigma_{n+R}\right]+
S\left[\rho_{n+R}^{(1)}\right]-
S\left[\rho_{n,n+R}^{(2)}\right].
\label{discord}
\ee
Here 
\be 
S\left[n|\sigma_{n+R}\right]=
\sum_{j=\pm1} p_j~ S\left[\frac{P_j\rho^{(2)}_{n,n+R}P_j}{p_j}\right],
\ee
$P_j=(1+j~\sigma_{n+R})/2$ is a projector on the measurement outcome $j=\pm1$ in the eigenbasis of a Pauli operator $\sigma_{n+R}$, and $p_j={\rm Tr}P_j\rho^{(2)}_{n,n+R}$ is a probability of this outcome. 
In the problem considered in this paper, the discord is symmetric, 
\be 
\delta_{n|n+R}=\delta_{n+R|n}\equiv\delta_R,
\ee 
and the minimum is achieved for $\sigma=\sigma^x$. Not surprisingly, the strongest ferromagnetic correlations are the most classical.

Like the mutual information, for slow enough $\tau_Q$ the discord becomes 
\bea
\delta_R &\approx&
\frac{\pi}{8} \left[ \frac{2 \pi}{\pi^2-4} - {\rm arctanh}\left(\frac{2}{\pi}\right) \right] 
\left(C^{xx}_R\right)^2 \nonumber\\
&=&
0.12 \left(C^{xx}_R\right)^2.
\eea
This suggests a space-time scaling,
\be 
\delta_R(t)=\hat\xi^{-4\Delta_x}F_\delta\left(t/\hat t,R/\hat\xi\right),
\label{discordscaling}
\ee
that is demonstrated by the collapsing plots in Fig. \ref{figdiscord}.

\begin{figure*}[t]
\includegraphics[width=1.95\columnwidth,clip=true]{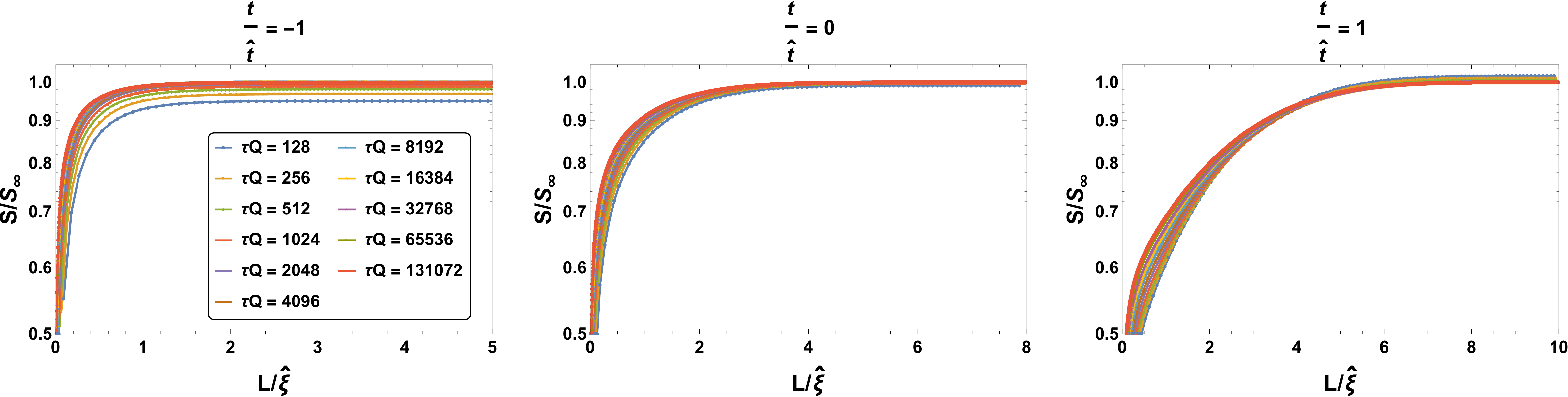}
\caption{  
Entanglement entropy in Eq. (\ref{S}).
The left, middle and right panels show the entropy before ($t/\hat t=-1$), at 
($t/\hat t=0$), and after ($t/\hat t=+1$) the critical point, respectively. 
Different colors of the plots correspond to different quench times $\tau_Q$.
All plots are rescaled: 
they are in function of the rescaled size of the block, $L/\hat\xi=L/\sqrt{\tau_Q}$, 
and the entropy is divided by 
$
S_\infty(t/\hat t)=
\frac{c}{3}\log k\hat\xi=
\frac{c}{3}\log k + \frac{c}{6}\log\tau_Q.
$
In consistency with the exact $c=\frac12$,
the best fits yield $c=0.524(3),0.5030(4),0.478(2)$ at $t/\hat t=-1,0,1$, respectively.
The plots in all three panels collapse asymptotically with increasing $\tau_Q$ demonstrating the space-time scaling (\ref{Sscaling}) for large enough $\tau_Q$.
The collapsed plots for large $\tau_Q$ are the scaling function
$F_S\left(t/\hat t,L/\hat\xi\right)$ in Eq. (\ref{Sscaling}).
}
\label{figentropy}
\end{figure*}
\begin{figure*}[t]
\includegraphics[width=1.95\columnwidth,clip=true]{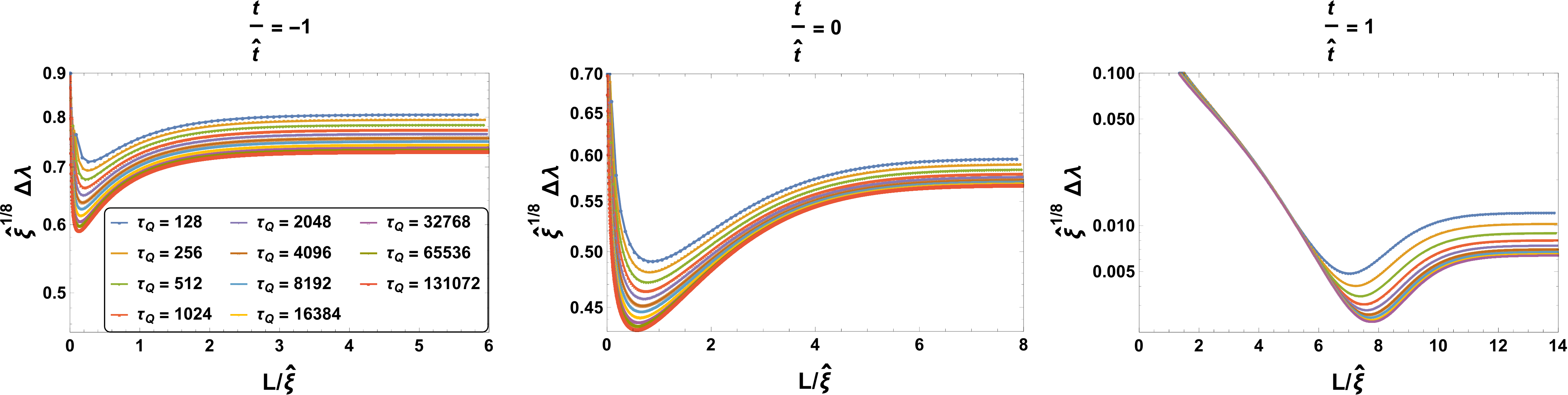}
\caption{  
Entanglement gap in Eq. (\ref{entgap}).
The left, middle and right panels show the gap before ($t/\hat t=-1$), at 
($t/\hat t=0$), and after ($t/\hat t=+1$) the critical point, respectively. 
Different colors of the plots correspond to different quench times $\tau_Q$.
All plots are rescaled: 
they are in function of the rescaled block size $L/\hat\xi=L/\sqrt{\tau_Q}$ and 
the gap is multiplied by $\hat\xi^{\beta/\nu}=\left(\sqrt{\tau_Q}\right)^{1/8}$.
The plots in all three panels collapse asymptotically with increasing $\tau_Q$ demonstrating the space-time
scaling (\ref{entgapscaling}) for large enough $\tau_Q$.
The collapsed plots for large $\tau_Q$ are the scaling function
$F_{\Delta\lambda}\left(t/\hat t,L/\hat\xi\right)$ in Eq. (\ref{entgapscaling}).
}
\label{figgap}
\end{figure*}
\subsection{ Scaling in block entropy }\label{sec:scalingentropy}

In order to go beyond the two-point correlations, one can consider a block of $L$ 
consecutive spins. Their reduced density matrix is obtained \cite{17} from a correlator matrix 
\be
\Pi_L~=~
\left(
\begin{array}{ccc}
A &,& B^{\dagger} \\  
B &,& 1-A  
\end{array}
\right)~,
\label{rhoblock}
\ee
where $A$ and $B$ are $L\times L$ Toeplitz matrices
$
A_{m,n}\equiv
\langle c_m c_n^\dagger \rangle= 
\alpha_{m-n} 
$
and 
$
B_{m,n}\equiv
\langle c_m c_n \rangle=
\beta_{m-n}
$.
The Hermitian $\Pi_L$ has eigenvalues $0\leq N_1\leq...\leq N_{2L}\leq1$ with a symmetry $N_m=1-N_{2L+1-m}$.
The eigenvalues $N_1,...,N_L$ are average occupation numbers for Bogoliubov quasiparticles $\Gamma_1,...,\Gamma_L$ localized on the $L$ sites of the block,
where we have a Bogoliubov transformation
\be 
c_n=\sum_{m=1}^L \left( U_{nm}\Gamma_m + V^*_{nm} \Gamma_m^\dag \right).
\ee
The $m$-th Bogoliubov mode $(U_{nm},V_{nm})$ is the eigenvector of $\Pi_L$ with the eigenvalue $N_m$.

In this Bogoliubov representation the reduced density matrix becomes a simple product
\be 
\rho_L=
\prod_{m=1}^L
\left[
N_m |1_m\rangle\langle1_m| +
(1-N_m) |0_m\rangle\langle0_m|
\right].
\ee
Here $|1_m\rangle$ ($|0_m\rangle$) is a state with one (zero) quasiparticle annihilated
by $\Gamma_m$. Consequently, the entropy of entanglement of the block of $L$ spins with the rest of the lattice is a sum
\bea 
S &=& -{\rm Tr}\rho_L\log\rho_L \nonumber\\
  &=& -\sum_{m=1}^L \left[ N_m\log N_m + (1-N_m)\log(1-N_m) \right] \nonumber\\
  &=& -\sum_{m=1}^{2L} N_m\log N_m .
\label{S}
\eea
Interestingly, the last sum is simply $-{\rm Tr}~\Pi_L\log\Pi_L$.

Near the critical point in the ground state with a long correlation length $\xi$, the entropy is $S=\frac{c}{3}\log \kappa\xi$ 
for a large block with $L\gg\xi$ and $S=\frac{c}{3}\log \kappa L$ for a relatively small one with $1\ll L\ll\xi$. Here $c=\frac12$ 
is the central charge and $\kappa\simeq 1$ a non-universal constant. With the KZ substitution $\xi\to\hat\xi$, motivated by 
the adiabatic-impulse approximation, in a dynamical transition we expect \cite{Cincio} respectively 
$S=\frac{c}{3}\log\kappa\hat\xi$ and $S=\frac{c}{3}\log\kappa L$. Beyond this approximation 
we allow $\kappa$ to be a function of the rescaled time $t/\hat t$. This argument suggests a space-time scaling
\be 
\frac{ S(t,L) }{ \frac{c}{3} \log \kappa(t/\hat t) ~ \hat\xi } =
F_S\left(t/\hat t,L/\hat\xi\right)
\label{Sscaling}
\ee
for large enough $\tau_Q$. Here we assume the normalization $F_S(t/\hat t,\infty)=1$ so that
the equation
\be 
S(t,\infty) = \frac{c}{3} \log \kappa(t/\hat t) \hat\xi \equiv S_{\infty}
\label{Sinfty}
\ee
defines implicitly the function $\kappa(t/\hat t)$. The scaling is demonstrated by the collapsing plots in Figure \ref{figentropy}. 
Since the entropy is only logarithmic in $\tau_Q$, the collapse requires much longer quench times than the spin-spin correlators.

\subsection{ Scaling in entanglement gap }\label{sec:scalinggap}

The entanglement gap is defined as a difference between two largest coefficients in the Schmidt decomposition 
between the block of $L$ spins and the rest of the spin chain or, equivalently, between square roots of the two largest 
eigenvalues of $\rho_L$. 
Since the largest eigenvalues are $(1-N_1)...(1-N_{L-1})(1-N_L)>(1-N_1)...(1-N_{L-1})N_L$,
the entanglement gap reads
\be 
\Delta\lambda=
\sqrt{(1-N_1)...(1-N_{L-1})}~\left(\sqrt{1-N_L}-\sqrt{N_L}\right).
\label{entgap}
\ee
According to Ref. \cite{EntGap}, 
in the ground state for a large block with $L\gg\hat\xi$ the entanglement gap should scale as
\be 
\Delta\lambda \sim \xi^{-\beta/\nu},
\ee
where $\beta=\frac18$ is the critical exponent for the order parameter. In a dynamical transition we substitute in the above 
formula $\xi$ with $\hat\xi$. Even more generally, for a finite block of size $L$, we can formulate a scaling law
\be 
\Delta\lambda = \hat\xi^{-\beta/\nu} ~ F_{\Delta\lambda}\left(t/\hat t,L/\hat\xi\right)
\label{entgapscaling}
\ee
expected to hold for slow enough $\tau_Q$. The collapsing plots in Figure \ref{figgap} demonstrate
this scaling law.

\section{ Conclusion }\label{sec:conclusion}
We made an extensive overview of the KZ space-time scaling in the quantum Ising chain.
We conclude that it is satisfied in the slow quench limit by all the quantities we have considered.
The limit is approached the fastest for the ferromagnetic correlator.
The scaling dimensions proved to be the same as in the static case.

It is tempting to speculate that our conclusion, while for the moment verified only in the Ising chain, 
may be a useful way of thinking about other quantum phase transitions as well as second order thermal phase 
transitions that cannot be probed with exactly solvable models. This would pave the way towards vast extension
of the renormalization from the static equilibrium critical phenomena to the space-time renormalization
of phase transition dynamics.
  
\acknowledgements
We appreciate discussions with  Andrew Daley, Bogdan Damski, Marek Rams, and Tommaso Roscilde.
This work was supported by the Polish National Science Center (NCN) under project DEC-2013/09/B/ST3/01603 (AF and JD), 
the Polish Ministry of Science and Higher Education under project Mobility Plus 1060/MOB/2013/0 (BG),
and Department of Energy under the Los Alamos National Laboratory LDRD Program (WHZ).

\appendix

\section{ Quasimomentum integrals in fermionic correlators }
\label{numalpha}

The fermionic correlators (\ref{alphaR},\ref{betaR}) are obtained by numerical integration in Mathematica.
In principle, the integrals should be done with the exact solutions (\ref{general}) in the full integration
range $k=0..\pi$, but it quickly becomes impractical above $\tau_Q\approx10$. Therefore we split the range
into two. For instance,
\bea
\beta_{R}= 
   \int_{0}^{A\hat k} \frac{dk}{\pi} u_kv_k^*\sin kR+
   \int_{A\hat k}^\pi \frac{dk}{\pi} u_kv_k^*\sin kR.
\label{betaRapp}
\eea
The first integral, covering more than the non-adiabatic regime $k=0...\hat k$ for $A>1$, is done exactly. 
In the second integral, where the evolution of the Bogoliubov modes is approximately adiabatic, 
we could approximate the Bogoliubov coefficients by just the positive-frequency adiabatic eigenstate:
\be 
\left(\begin{array}{c} u_k\\ v_k\end{array}\right)\approx
\left(\begin{array}{c} U_k\\ V_k\end{array}\right),
\ee
compare Eq. (\ref{stBdG}).
However, a much better approximation is obtained at very little expense by including also a first order 
perturbative correction:
\be 
\left(\begin{array}{c} u_k\\ v_k\end{array}\right)\approx
\frac{1}{\sqrt{1+|B|^2}}
\left[
\left(\begin{array}{c} U_k\\ V_k\end{array}\right)+
B
\left(\begin{array}{c} -V_k\\ U_k\end{array}\right)
\right].
\ee
Here $B$ is an amplitude of excitation to the adiabatic negative-frequency mode,
\be 
B=
\frac{e^{-2i\varphi}}{2\omega_k(g)\tau_Q}
\left(\begin{array}{c} U_k\\ V_k\end{array}\right)^\dag
\frac{d}{dg}
\left(\begin{array}{c} -V_k\\ U_k\end{array}\right).
\ee
The phase $\varphi$ drops out in Eq. (\ref{betaRapp}). The results do not depend on $A$ in the range $2..3$.



\end{document}